\documentstyle[prl,twocolumn,aps,english]{revtex}

\def\beq{\begin{equation}}
\def\eeq{\end{equation}}
\def\6{\langle}
\def\9{\rangle}
\def\DD{{\cal D{}}}
\def\ptp{\psi\theta\phi}
\def\xyz{\xi\eta\zeta}
\def\mnr{\mu\nu\rho}
\def\abc{\alpha\beta\gamma}

\begin{document}
\draft

\title{Transmission of a Cartesian Frame by a Quantum System}

\author{Asher Peres and Petra F. Scudo}
\address{Department of Physics, Technion---Israel Institute of
Technology, 32000 Haifa, Israel}

\maketitle
\begin{abstract}
A single quantum system, such as a hydrogen atom, can transmit a
Cartesian coordinate frame (three axes). For this it has to be prepared
in a superposition of states belonging to different irreducible
representations of the rotation group. The algorithm for decoding such
a state is presented, and the fidelity of transmission is evaluated.
\end{abstract}

\bigskip

There has recently been considerable progress in devising ways to
indicate a spatial direction by means of quantum particles.  This type of
information cannot be represented by a sequence of symbols like 0 and~1,
unless the emitter (Alice) and the receiver (Bob) have prearranged a
common coordinate system for specifying the numerical values of relevant
angles. Physical objects have to be sent. Preceding works [1--4] have
considered the use of spins for transmitting a single direction. The
simplest method~\cite{mp} is to send these spins polarized along
the direction that one wishes to indicate. This, however, is not
the most efficient procedure: when two spins are transmitted, a
higher accuracy is achieved by preparing them with {\it opposite\/}
polarizations~\cite{gp}. If there are more than two spins, optimal
results are obtained with {\it entangled\/} states \cite{bbbmt,petra}.

This Letter presents a method for the transmission of a complete
Cartesian frame. If many spins are available, a simple possibility
would be for Alice to use half of them for indicating her $x$ axis and
the other half for her $y$ axis. However, the two directions found
by Bob may not then be exactly perpendicular, because of quantum
``uncertainties.'' Some adjustment will be needed to obtain the best
estimates for the $x$ and $y$ axes before Bob can infer from them his
guess of Alice's $z$ direction. This method is not optimal, and it is
obviously not possible to proceed in this way if a {\it single\/} quantum
messenger is available. Here we shall show how a single hydrogen atom
(formally, a spinless particle in a Coulomb potential) can transmit a
complete frame.

Consider the $n$-th energy level of that atom (a Rydberg state). Its
degeneracy is $d=n^2$ because the total angular momentum may take
values $j=0,\cdots,n-1$, and for each one of them $m=-j,\cdots,j$. Alice
indicates her $xyz$ axes by sending the atom in a state

\beq |A\9=\sum_{j=0}^{n-1}\;\sum_{m=-j}^j a_{jm}\,|j,m\9,\label{A}\eeq
with normalized coefficients $a_{jm}$ that will be specified below. Bob
then performs a covariant measurement~\cite{holevo} in order to evaluate
the Euler angles $\ptp$ that would rotate his own $xyz$ axes into a
position parallel to Alice's axes. Bob's detectors (ideally, there is
an infinite number of them~\cite{finite}) have labels $\ptp$ and the
mathematical representation of his apparatus is a {\it positive operator
valued measure\/} (POVM)~\cite{qt,maxwell}, namely a resolution of
identity by a set of positive operators:

\beq  \int d_{\ptp}\,E(\ptp)={\bf1}, \label{povm}\eeq
where $d_{\ptp}\equiv\sin\theta d\psi d\theta d\phi/8\pi^2$ is the
$SO(3)$ Haar measure for Euler angles~\cite{edm}, and
$E(\ptp)=|\ptp\9\6\ptp|$. The vectors $|\ptp\9$ will be specified
below. The probability that the detector labelled $\ptp$ is excited is
given by

\beq P(\ptp)= \6A|d_{\ptp} E(\ptp)|A\9=
  d_{\ptp}\,|\6A|\ptp\9|^2. \label{prob} \eeq
Our task is to construct vectors $|\ptp\9$ such that Eq.~(\ref{povm})
is satisfied (that is, the probabilities sum up to one) and Bob's
expected error is minimal.

Following the method of Ref.~\cite{petra}, we define a fiducial vector
for Bob,

\beq |B\9=\sum_{j=0}^{n-1}\sqrt{2j+1}\sum_{m=-j}^j b_{jm}\,|j,m\9,
  \label{B} \eeq
where the coefficients $b_{jm}$ are normalized for each $j$ separately:

\beq \sum_{m=-j}^j\,|b_{jm}|^2=1\qquad\forall j. \label{b} \eeq
In Ref.~\cite{petra} a single value of $m$ was used; here we need all the
values. Note that Eq.~(\ref{A}) was written with Alice's notations ($m$
is the angular momentum along her $z$ axis), while Eq.~(\ref{B}) is
written with Bob's notations ($m$ refers to his $z$ axis). This issue
will be dealt with later.

We now define

\beq |\ptp\9=U(\ptp)\,|B\9, \eeq
where $U(\ptp)$ is the unitary operator for a rotation by Euler angles
$\ptp$. Note that since $|B\9$ is a direct sum of vectors, one for each
value of $j$, then likewise $U(\ptp)$ is a direct sum with one term
for each irreducible representation,

\beq U(\ptp)=\sum_j\oplus\DD^{(j)}(\ptp), \eeq
where the $\DD^{(j)}(\ptp)$ are the usual irreducible unitary rotation
matrices~\cite{edm}.  To prove that Eq.~(\ref{povm}) is satified, we
note that its left hand side is invariant if multiplied by $U(\mnr)$
on the left and $U(\mnr)^\dagger$ on the right, for any arbitrary Euler
angles $\mnr$ (because these unitary matrices represent group elements
and therefore have the group multiplication properties)~\cite{perel}. It
then follows from a generalization of Schur's lemma~\cite{wigner} that
the left hand side of (\ref{povm}) is a direct sum of {\it unit\/}
matrices, owing to the presence of the factor $(2j+1)$ which is the
dimensionality of the corresponding irreducible representation. Therefore
Eq.~(\ref{povm}) is satisfied.

The detection probability (\ref{prob}) can thus be written as $P(\ptp)=
d_{\ptp}\,|\6A|U(\ptp)|B\9|^2.$ To compute this expression explicitly,
we must use a uniform system of notations for $|A\9$ and $|B\9$ ---
recall that Eq.~(\ref{A}) was written in Alice's basis, and
Eq.~(\ref{B}) in Bob's basis. It is easier to rewrite Alice's vector
$|A\9$ in Bob's language. For this we have to introduce the Euler angles
$\xyz$ that rotate Bob's $xyz$ axes into Alice's axes (that is, $\xyz$
are the true, but unknown values of the angles $\ptp$ sought by Bob).
The unitary matrix $U(\xyz)$ represents an {\it active\/} transformation
of Bob's vectors into Alice's. Therefore, $U(\xyz)^\dagger$ is the {\it
passive\/} transformation \cite[p.~216]{qt} from Bob's notations to
those of Alice, and $U(\xyz)$ is the corresponding transformation
from Alice's notations to Bob's. Written in Bob's notations, Alice's
vector $|A\9$ becomes $U(\xyz)|A\9$ so that, in Eq.~(\ref{prob}), $\6A|$
becomes $\6A|U(\xyz)^\dagger$. Let us therefore define

\beq U(\abc)=U(\xyz)^\dagger\,U(\ptp). \eeq
The Euler angles $\abc$ have the effect of rotating Bob's Cartesian
frame into his {\it estimate\/} of Alice's frame, and then rotating
back the result by the {\it true\/} rotation from Alice's to Bob's
frame. That is, the angles $\abc$ indicate Bob's measurement error,
and the probability of that error is

\beq P(\abc)= d_{\abc}\,|\6A|U(\abc)|B\9|^2, \label{errprob}\eeq
where $d_{\abc}=\sin\beta d\alpha d\beta d\gamma/8\pi^2$. Note that
in the above equation $|A\9$ is written with Alice's notations as in
(\ref{A}), and $|B\9$ with Bob's notations as in (\ref{B}). 

Of course Bob cannot know the values of $\abc$. His measurement only
yields some value for $\ptp$. The following calculation that employs
$\abc$ has the sole purpose of estimating the expected accuracy of the
transmission (which does not depend on the result $\ptp$).

We must now choose a suitable quantitative criterion for that accuracy.
When a single direction is considered, it is convenient to define the
error~\cite{helstrom} as $\sin(\omega/2)$, where $\omega$ is the angle
between the true direction and the one estimated by Bob. The mean square
error is

\beq \6\sin^2(\omega/2)\9=(1+\6\cos\omega\9)/2=1-F, \eeq
where $F$ is usually called the {\it fidelity\/}~\cite{petra}. When
we consider a Cartesian frame, we likewise define fidelities for
each axis. Note that $\cos\omega_k$ (for the $k$-th axis) is given
by the corresponding diagonal element of the orthogonal (classical)
rotation matrix.  Explicitly, we have~\cite{goldstein}

\beq \cos\omega_z=\cos\beta, \eeq and
\beq \cos\omega_x+\cos\omega_y=(1+\cos\beta)\,\cos(\alpha+\gamma), 
 \label{xy} \eeq
whence, by Euler's theorem,

\beq \cos\omega_x+\cos\omega_y+\cos\omega_z=1+2\cos{\mit\Omega},
 \label{xyz} \eeq
where $\mit\Omega$ has a simple physical meaning: it is the angle for
carrying one frame into the other by a single rotation.

The expectation values of these expressions are obtained from
Eq.~(\ref{errprob}):

\beq \6f(\abc)\9=\int d_{\abc}\,|\6A|U(\abc)|B\9|^2\,f(\abc), \eeq
where, explicitly,

\beq \6A|U(\abc)|B\9=\sum_{j,m,r}\;a_{jm}^*\,b_{jr}
  \6j,m|\DD^{(j)}(\abc)|j,r\9. \eeq
The unitary irreducible rotation matrices $\DD^{(j)}(\abc)$ have
components~\cite{edm}
\beq \6j,m|\DD^{(j)}(\abc)|j,r\9=
  e^{i(m\alpha+r\gamma)}\,d^{(j)}_{mr}(\beta), \eeq
where the $d^{(j)}_{mr}(\beta)$ can be expressed in terms of Jacobi
polynomials. Collecting all these terms, we finally obtain, after many
tedious analytical integrations over products of Jacobi
polynomials~\cite{rain},

\beq \6f(\abc)\9=\sum f_{jkmnrs}\,a_{jm}^*\,b_{jr}\,a_{kn}\,b_{ks}^*,
 \label{fabab} \eeq
where the numerical coefficients $f_{jkmnrs}$ depend on our choice of
$f(\abc)$. The problem is to optimize the components $a_{jm}$
(normalized to 1), and $b_{jm}$ satisfying the constaints (\ref{b}),
so as to maximize the above expression. For further use, it is convenient
to define a matrix

\beq M_{jm,kn}=\sum_{r,s}f_{jkmnrs}\,b_{jr}\,b^*_{ks},\label{bilin}\eeq
so that 

\beq \6f(\abc)\9=\sum M_{jm,kn}\,a^*_{jm}\,a_{kn}=\6A|M|A\9.
 \label{Maa} \eeq

First consider a simple case: to transfer only the $z$ axis, we wish to
maximize $\6\cos\beta\9$. An explicit calculation yields

\beq f_{jkmnrs}=\delta_{mn}\,\delta_{rs}\,g_{jk}.\label{fddg} \eeq 
The matrix $g_{jk}$ which is defined by the above equation has
nonvanishing elements

\beq g_{jj}=ns/[j(j+1)], \eeq  and
\beq g_{j,j-1}=g_{j-1,j}={1\over j}
 \sqrt{{(j^2-n^2)(j^2-s^2)\over4j^2-1}}. \eeq
The $\delta_{mn}$ term in (\ref{fddg}) implies that, for any choice of
Bob's fiducial vector $|B\9$, the matrix $M$ in (\ref{Maa}) is
block-diagonal, with one block for each value of $m$. The optimization
of Alice's signal results from the highest eigenvalue of that matrix.
This is the highest eigenvalue of one of the blocks, so that a single
value of $m$ is actually needed. A similar (slightly more complicated)
argument applies if Alice's vector is given and we optimize Bob's
fiducial vector. This result proves the correctness of the intuitive
assumption that was made in \cite{bbbmt,petra} where a single value of
$m$ was used. It was then found that when $m=0$ (the optimal value)
the expected error asymptotically behaves as $1.446/d$, where $d$ is
the effective number of Hilbert space dimensions. In the present case,
$d=(j_{{\rm max}}+1)^2$, which is the degeneracy of the \mbox{$n$-th}
energy level.

If we want to transfer two axes, we use Eq.~(\ref{xy}) and calculate
the matrix elements for $\6(1+\cos\beta)\cos{(\alpha+\gamma)}\9$. (It is
curious that they are simpler than those for $\6\cos{(\alpha+\gamma)}\9$
alone.) We obtain

\beq f_{jkmnrs}=\delta_{m,n-1}\,\delta_{r,s-1}\,h_{jk}+
 \delta_{n,m-1}\,\delta_{s,r-1}\,h_{kj}, \label{h} \eeq
where the nonvanishing elements of $h_{jk}$ are

\begin{eqnarray}
h_{jj}&=&\frac{[(j-n+1)(j+n)(j-s+1)(j+s)]^{1/2}}{2j(j+1)}, \\
h_{j,j-1}&=&\frac{[(j-n+1)(j-n)(j-s+1)(j-s)]^{1/2}}
 {2j(4j^2-1)^{1/2}}, \\
h_{j-1,j}&=&\frac{[(j+n-1)(j+n)(j+s-1)(j+s)]^{1/2}}
 {2j(4j^2-1)^{1/2}}. \end{eqnarray}
Note that the $h_{jk}$ matrix, whose elements depend on $n$ and $s$,
is not symmetric (while $g_{jk}$ was). This is because it comes from
the operator $e^{i(\alpha+\gamma)}$ which is not Hermitian. However
the two terms of (\ref{h}) together, which corresponds to
$\cos(\alpha+\gamma)$, have all the symmetries required by the other
terms $a_{jm}^*\,b_{jr}\,a_{kn}\,b_{ks}^*$ in Eq.~(\ref{fabab}).
Finally, if we wish to optimize directly the three Cartesian axes
(without losing accuracy by inferring $z$ from the approximate knowledge
of $x$ and $y$) we use all the terms of (\ref{xyz}), that is, both
those of (\ref{fddg}) and of (\ref{h}).

It now remains to find the vectors $|A\9$ and $|B\9$ that minimize
the transmission error. For small values of $j$, we used Powell's
method~\cite{nr} without imposing any restrictions on $|A\9$ and $|B\9$
other than their normalization conditions. As intuitively expected,
we found that the optimal vectors satisfy

\beq b_{jm}=a_{jm}\left(\sum_n|a_{jn}|^2\right)^{-1/2},
  \qquad\forall j. \eeq
This means that Bob's vector should look as much as possible as Alice's
signal, subject to the restrictions imposed by the constraint~(\ref{b}).

Taking this property for granted is the key to a more efficient
optimization method, as follows: assume any $b_{jm}$, so that the
bilinear form (\ref{bilin}) is known. Find its highest eigenvalue and the
corresponding eigenvector $a_{jm}$. From the latter, get new components
$b_{jm}$ by means of (\theequation), and repeat the process until it
converges (actually, a few iterations are enough). The results are shown
in Fig.~1. It is seen that there is little advantage in optimizing only
two axes, if for any reason the third axis is deemed less important.
If the three axes are simultaneously optimized, the mean square error
(per axis) asymptotically tends to $3.168\,d^{-0.586}$.

It is not surprising that this result is weaker than the one for a
single axis, which was $1.446/d$. The obvious reason is that we are now
transmitting a three-dimensional rotation operation that can be applied
to any number of directions, not only to three orthogonal axes. Indeed,
consider any set of unit vectors ${\bf e}^\mu_m$, where $m=1,2,3$, and
$\mu$ is a label for identifying the vectors. Let $w_\mu$ be a positive
weight factor attached to each vector, indicating its importance. Let
$R(\abc)$ be the classical orthogonal rotation matrix~\cite{goldstein}
for Euler angles $\abc$.  Then the cosine of the angle between Bob's
estimate of ${\bf e}^\mu_m$ and the true direction of that vector is

\beq \cos\omega_\mu=
 \sum_{m,n}R_{mn}(\abc)\,{\bf e}^\mu_m\,{\bf e}^\mu_n. \eeq
With the same notations as before, we have

\beq \6f(\abc)\9=\sum_\mu w_\mu\,\6\cos\omega_\mu\9=
 \sum_{m,n}\6R_{mn}(\abc)\9\,C_{mn}, \eeq
where

\beq C_{mn}=\sum_\mu w_\mu\,{\bf e}^\mu_m\,{\bf e}^\mu_n. \eeq
This is a positive matrix which depends only on the geometry of the
set of vectors whose transmission is requested. We can now diagonalize
$C_{mn}$ and write it in terms of three orthogonal vectors, possibly
with different weights. Therefore, no essentially new features follow
from considering more than three directions.

Finally, we note that all the above calculations, as well as those in
preceding works [1--4], assume that Alice and Bob have coordinate frames
with the same chirality (this can be checked locally by using weak
interactions). If the chiralities are opposite, then all the directions
inferred by Bob should be reversed (because directions are polar vectors
while spins are axial vectors).

In summary, we have shown that a single structureless quantum system (a
point mass in a Coulomb potential) can transfer information on the
orientation of a Cartesian coordinate system with arbitrary accuracy.
No classical carrier would be able to achieve this result, unless it has
an asymmetric internal structure. This is one more example of the
remarkable ability of quantum systems to encode information more
efficiently than classical ones.

\bigskip We thank Ramon Mu\~noz-Tapia for helpful comments. Work by AP
was supported by the Gerard Swope Fund and the Fund for Encouragement
of Research. PFS was supported by a grant from the Technion Graduate
School.

\vfill

FIG. 1. \ Mean square error (per axis) for the transmission of the
directions of one, two, or three axes, by a single quantum carrier.

\begin{thebibliography}{9}
\bibitem{maxwell}Owing to Gleason's theorem, any detection procedure
is mathematically represented by a POVM.  However, quantum mechanics is
only a theoretical framework and it does not tell how to actually build
the corresponding experimental setup. Likewise, Maxwell's equations do
not say how to build an electric motor, and thermodynamics does not tell
how to build steam engines.
\end{thebibliography}
\end{document}